\documentclass[
prb,twocolumn,superscriptaddress]{revtex4-2}
\usepackage{graphicx}
\usepackage{color,bm}
\usepackage[colorlinks,bookmarks=false,citecolor=blue, urlcolor=blue]{hyperref}
\usepackage{amsmath,amssymb}
\usepackage{mathrsfs}
\usepackage{float}
\usepackage{siunitx}
\usepackage[export]{adjustbox} %to align image to left or right within figure
\usepackage{multirow}
\usepackage{svg}
\usepackage{array} %vertically aligns content in rows
\usepackage{colortbl}%colored hline in table
\providecommand\vect[1]{\boldsymbol{#1}}

\begin{document}

\title{A Construction Framework for Three-Dimensional Topological Magnetic Textures}

\author{
Maria Azhar$^{1,*}$, Sandra Chulliparambil Shaju$^{1,*}$ and Karin Everschor-Sitte}

\address{Faculty of Physics and Center for Nanointegration Duisburg-Essen (CENIDE), University of Duisburg-Essen, 47057 Duisburg, Germany\\
$^{*}$ These authors contributed equally.}

%\subject{Physics}


%\corres{Karin Everschor-Sitte\\
%\email{karin.everschor-sitte@uni-due.de}}

\begin{abstract}
Three-dimensional magnetic systems offer diverse possibilities for constructing complex topological textures from simpler magnetic configurations. Here, we present different routes for constructing complex three-dimensional magnetic textures from simpler parent configurations and discuss their impact on the topology of the resulting structures. We identify four construction principles: 
(i) adding transverse modulations, 
(ii) adding structures, 
(iii) entangling textures (e.g.
braiding, or twisting), 
and (iv)
introducing singularities.
 Using the skyrmion tube as an explicit parent configuration, we show how these principles give rise to several known three-dimensional magnetic textures. We then apply the same principles to predict novel textures derived from a meron tube, which are confirmed by micromagnetic simulations to be metastable in chiral magnets. 
Our results establish general principles for the systematic construction and discovery of complex three-dimensional magnetic textures.
\end{abstract}

\maketitle

%\tableofcontents

\section{Introduction}
Topological magnetic textures emerge from the interplay of competing magnetic interactions. In two dimensions, this interplay gives rise to well-established structures such as skyrmions~\cite{Bogdanov1989, Muhlbauer2009, Yu2010, schulz2012, Everschor2018, sabri2026} and merons~\cite{Usov1993, cowburn1999, shinjo2000}.
 In three dimensions (3D), the additional spatial degree of freedom allows for much richer textures~\cite{donnelly2017,fullerton2026,Azhar2026}, including skyrmion tubes~\cite{Muhlbauer2009, zhang2021dynamic, birch2020, Milde2013, yu2020, Seki2022}, torons~\cite{Liu2018, Leonov2018, Muller2020, hu2021, li2022}, screw dislocations~\cite{Azhar2022, Zheng2023}, and Hopfions~\cite{Liu2018, Sutcliffe2017, sutcliffe2018, voinescu2020, liu2020, Kent2021, Zheng2023, liu2024, Shaju2026, kasai2026hopfion}.

Despite their apparent diversity, many 3D textures can be viewed as modifications of simpler precursor configurations, which we refer to as ``parent'' configurations in the following. In this work, we introduce this perspective as a systematic framework for constructing complex textures.
 Starting from a generic parent configuration, we identify four basic construction principles. 

We show that these principles and their combination 
give rise to different ``families'' of magnetic textures
 and that they provide a systematic way of relating their topology to the underlying construction.

Beyond establishing the general framework, we first illustrate its application using the skyrmion tube as a parent configuration. Applying the four construction principles gives rise to a range of 3D textures previously reported in the literature, providing a unified view of structures that are often considered separately. We then extend the same approach to a meron tube, from which we predict several new 3D textures. Micromagnetic simulations show that these configurations can occur as metastable states in chiral magnets. 
Thus, the construction principles provide not only a framework for organizing known textures, but also a systematic approach for predicting and discovering new three-dimensional magnetic structures.

We now introduce these principles in a general form and discuss how they can be applied to different parent configurations.

\begin{figure}
    \centering
    \includegraphics[width=\linewidth]{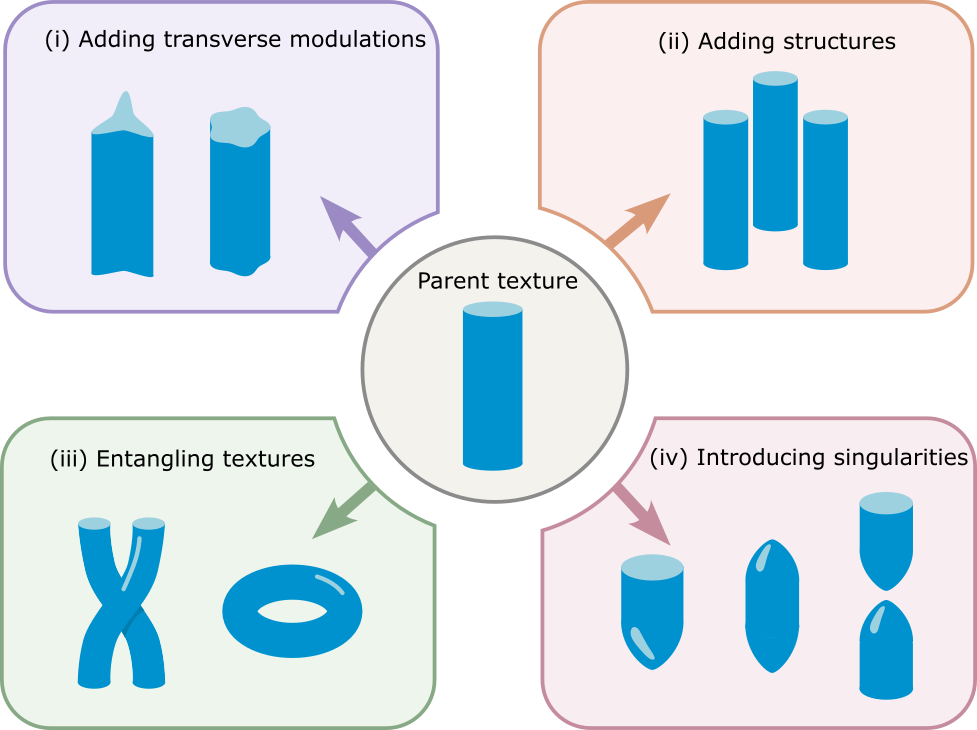}
    \caption{
    Construction principles for generating complex 3D magnetic textures from a simpler parent texture: (i) adding transverse modulations, (ii) adding structures, (iii) entangling textures, e.g.\ by braiding or connecting parent structures, and (iv) introducing singularities. 
    }
    \label{fig:summary_routes}
\end{figure}

\section{General construction framework for 3D magnetic textures}
\label{sec:general}

We consider different routes for constructing complex 3D magnetic textures from simpler parent configurations, as summarized in Fig.~\ref{fig:summary_routes}.

In general, a parent texture can be modified~\footnote{For a generic tubular parent texture, we use a coordinate $s$ along the tube axis and transverse coordinates $\mathbf r_\perp=(u,v)$ in the local perpendicular plane. Variations along $s$ are termed longitudinal, while variations in $\mathbf r_\perp$ are transverse. In each cross-section, the in-plane magnetization winds with \textit{vorticity} $V$, while the \textit{helicity} $\chi$ sets the angular phase of this winding. The \textit{polarity} specifies the orientation of the magnetization at the tube core relative to the surrounding background. A variation $\chi=\chi(s)$ therefore produces a longitudinal \textit{twist}, whereas transverse modulations can change the vorticity. Thus, helicity twisting and vorticity changes represent longitudinal and transverse modifications of the parent tube, respectively.} according to four construction principles:
\begin{enumerate}
    \item \textbf{Adding transverse modulations}: Changing the vorticity, thereby modifying the internal structure of the parent texture.
    \item \textbf{Adding structures}: Introducing additional topological structures within or around the parent texture, resulting in composite configurations.
    \item \textbf{Entangling textures}: Connecting or braiding one or more parent textures to generate more complex 3D configurations. 
    \item \textbf{Introducing singularities}: Introducing Bloch points or other singularities that terminate, connect, or restructure parts of the parent texture.
\end{enumerate}
These principles are not mutually exclusive and can be combined to generate further families of three-dimensional magnetic textures.

In the following, we illustrate the framework using two parent configurations. We first show that several previously reported 3D textures can be derived from the same skyrmion-tube parent. 
We then apply the construction principles to a meron tube to predict new textures and demonstrate their metastability through micromagnetic simulations in a representative chiral-magnet model.

\subsection{Skyrmion tube as a parent configuration}

\begin{figure}[b!]
    \centering
    \includegraphics[width=\linewidth]{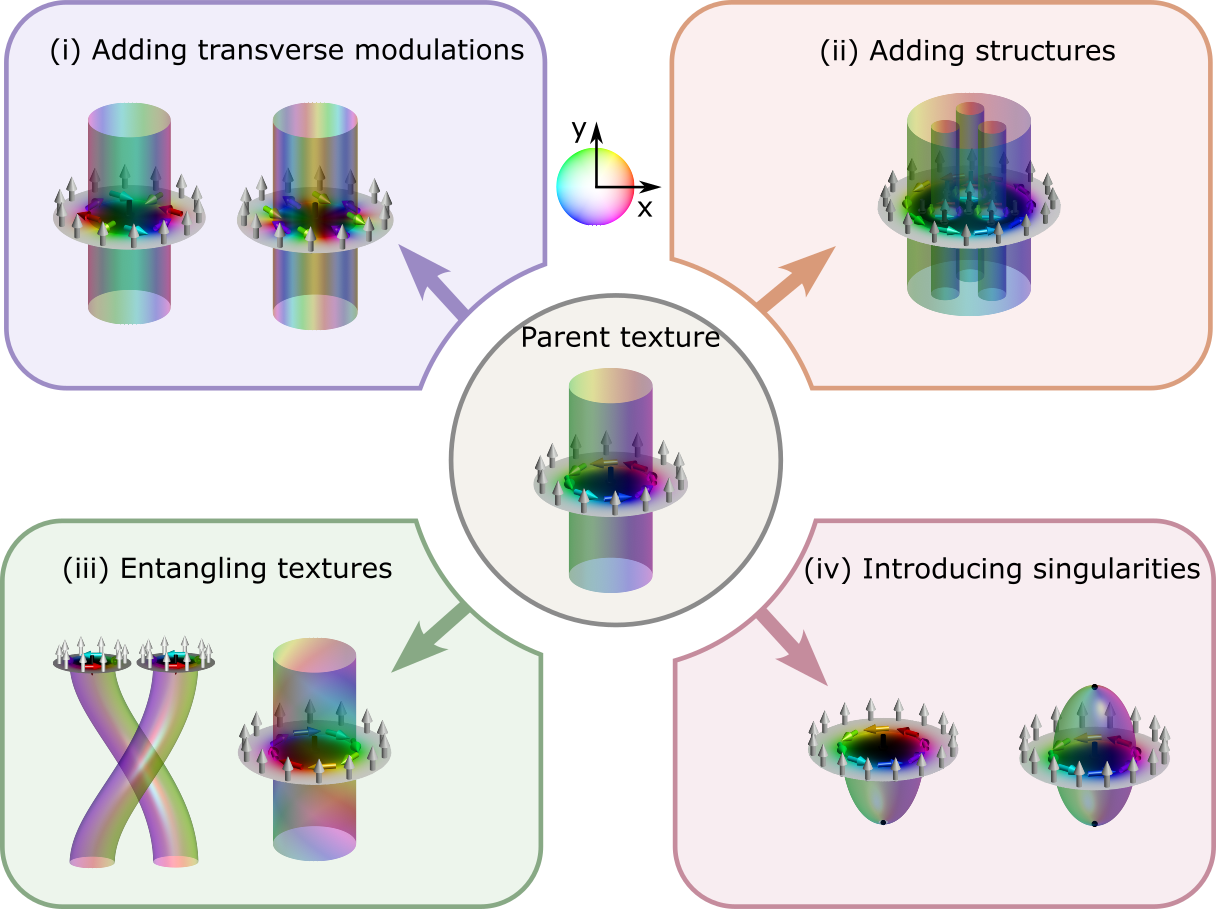}
    \caption{Overview of selected topological textures constructed from a skyrmion tube (center) according to four construction principles: adding vorticity modulations (upper left), adding skyrmion tubes (upper right), entangling skyrmion tubes (lower left), and introducing Bloch points (lower right). The textures are represented by their $m_z=0$ isosurfaces, with transparency adjusted for clarity. Circular planes show representative cross-sections, with arrows indicating the local magnetization direction. The isosurfaces and arrows are colored according to the colormap shown in the inset.}
    \label{fig:skyrmion_parent}
\end{figure}

We begin with a skyrmion tube~\cite{Muhlbauer2009, Bogdanov1989, zhang2021dynamic, birch2020, Milde2013, yu2020, Seki2022} as a parent configuration and illustrate how the four construction principles introduced above lead to a range of known 3D magnetic textures, see Fig.~\ref{fig:skyrmion_parent}.

Following route (i), additional in-plane modulations can be introduced into the skyrmion tube, modifying its internal structure and, in particular, its vorticity, which characterizes the winding of the in-plane magnetization around the skyrmion core. Increasing this winding gives rise to higher-order skyrmionic textures with higher vorticity and, correspondingly, higher topological charge~\cite{rodrigues2018, zhang2016, shustin2023, hassan2024, koraltan2026, titze2026}.

Following route (ii), \textit{adding structures}, additional skyrmion tubes can be embedded within or around the parent tube. This gives rise to composite structures such as skyrmioniums~\cite{bogdanov1999, Zhang2018, zhang2016control, msiska2022}, target skyrmion tubes~\cite{Kent2021}, and skyrmion sacks~\cite{rybakov2019} . 

Route (iii), \textit{entangling textures}, encompasses twisting, braiding, and closing skyrmion tubes. A helicity variation 
\footnote{A variation of the helicity $\chi=\chi(s)$.} 
along the tube axis produces twisted skyrmion tubes~\cite{saji2025, Kasai2026, Shaju2026}, while multiple tubes can intertwine to form braided skyrmion structures~\cite{zheng2021, knapman2024}.

Finally, route (iv), \textit{introducing singularities}, involves the incorporation of point defects in the magnetization field, commonly referred to as Bloch points or emergent magnetic monopoles. Such singularities allow the skyrmion number of successive cross sections to change along the tube. 
For example, a single Bloch point can terminate a skyrmion tube, giving rise to a chiral bobber~\cite{Rybakov2015, Zheng2018, Ran2021, ahmed2018, mankenberg2025}.
In contrast, a pair of oppositely charged Bloch points can confine a finite skyrmion-string segment, giving rise to textures such as torons~\cite{Liu2018, Leonov2018, Muller2020, hu2021, li2022} and skyrmion cocoons~\cite{grelier2022, grelier2023}. 

Combining different construction principles provides a route to more complex textures. For example, twisting and bending a single skyrmion tube and joining its ends appropriately can produce a closed three-dimensional texture, such as a hopfion~\cite{Liu2018, Sutcliffe2017, sutcliffe2018, voinescu2020, liu2020, Kent2021, Zheng2023, liu2024, Shaju2026, kasai2026hopfion}. Singularities can further mediate the splitting or merging of skyrmion tubes, yielding branched or Y-shaped configurations~\cite{Milde2013}
Another example is provided by hybrid skyrmion--antiskyrmion strings connected through a Bloch-point quadrupole~\cite{Yasin2024}.

\subsection{Meron tube as a parent configuration}
\label{sec:meron}

In the following, we consider a meron tube~\footnote{A meron tube is a tubular texture whose transverse cross-section is a half-skyrmion.} as the parent configuration.
Starting from this parent state, we construct a family of three-dimensional textures by combining several elementary transformation routes. In what follows, we distinguish between meron tubes with translationally invariant cores, which serve as the parent configurations, and twisted meron tubes, obtained by applying route (iii).

\subsubsection{Meron tube with translationally invariant core}
\begin{figure}
    \centering
    \includegraphics[width=\linewidth]{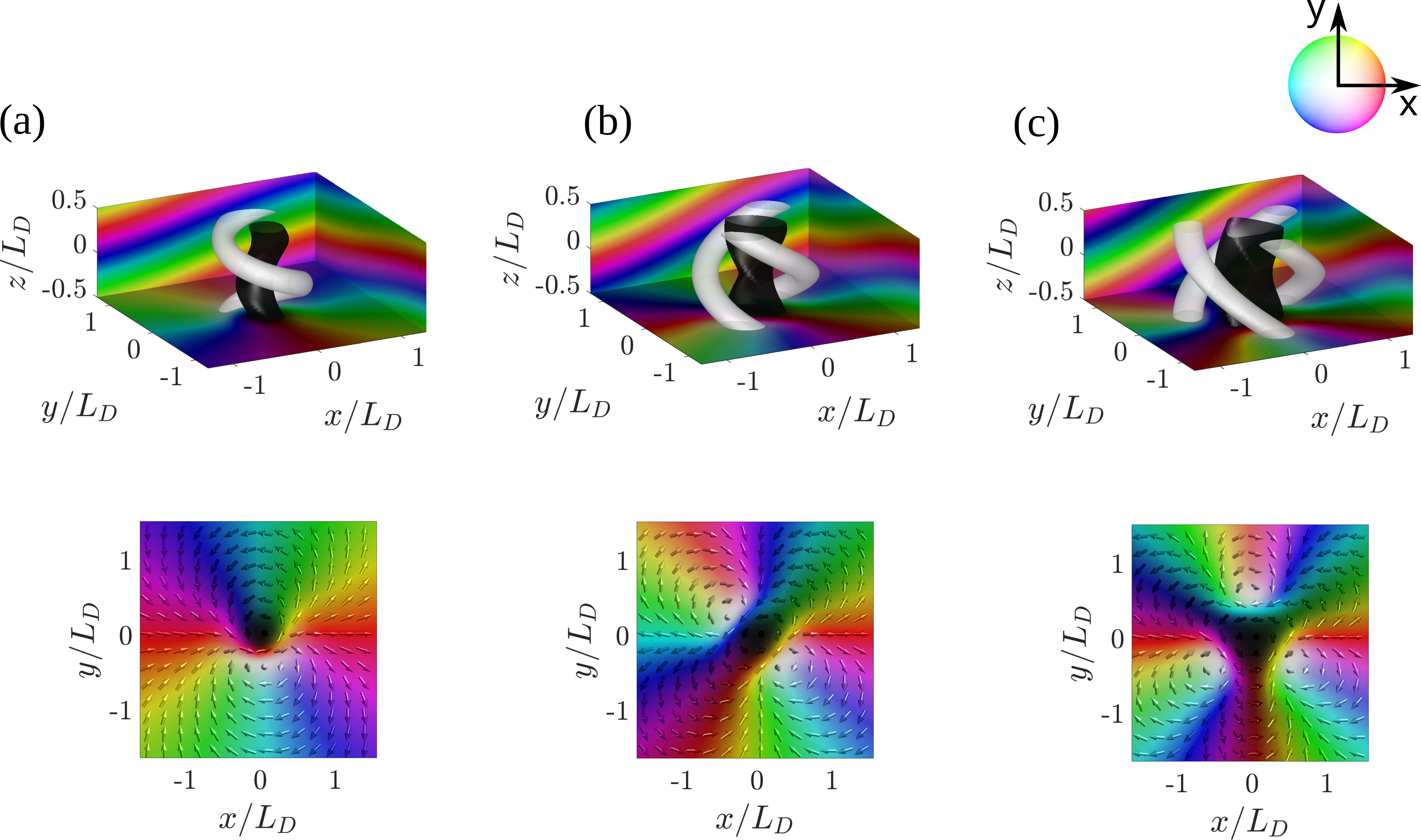}
    \caption{Family of magnetic textures derived from a meron tube by combining routes  (ii)--(iii). The panels differ in the number of additional textures arranged around the central meron tube. The structures are visualized using the $m_z=\pm0.75$ isosurfaces, while the background colors represent the magnetization in the $xz$, $yz$, and $xy$ planes. Bottom row: corresponding $xy$ cross-sections at $z=0$ including arrows representing the local magnetization direction; see inset for the color code. The same style is used throughout the following figures.  
    }
    \label{fig:braided_textures}
\end{figure}

Here we present three different types of families that can be constructed from meron tube with a translationally invariant core along the tube axis, shown in Figs.~\ref{fig:braided_textures} and ~\ref{fig:half_hopfion}.

The family of textures shown in Fig.~\ref{fig:braided_textures} can be constructed from a meron tube by combining routes (ii)--(iii). Starting from the central meron tube, additional meron tubes are incorporated following route (ii), \textit{adding structures}. Then, using route (iii), \textit{entangling textures}, these additional meron tubes are braided around the central meron tube, giving rise to the resulting three-dimensional textures.

Another interesting two-meron texture is obtained by combining routes (iii) and (ii). 
First, an additional meron tube is introduced following route (ii). Following route (iii), \textit{entangling textures}, this tube is then twisted and closed into a toroidal structure surrounding the central parent meron tube. The resulting texture, referred to as a twiston configuration~\cite{Azhar2022, Azhar2026}, is depicted and further discussed in 
Sec.~\ref{sec:general}\ref{sec:composite}.

\subsubsection{Twisted meron tube}
\begin{figure}
\centering
\includegraphics[width=\linewidth]{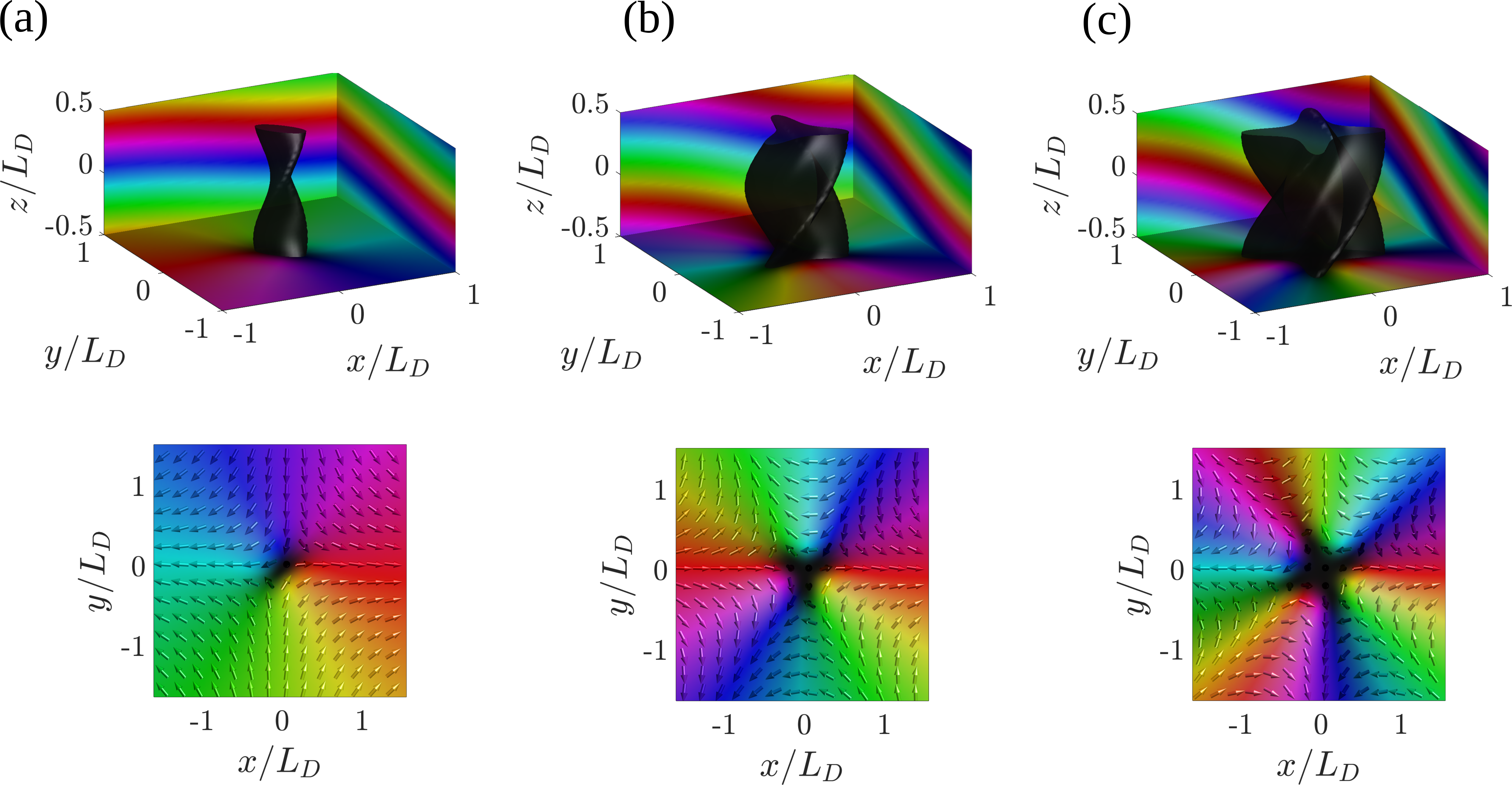}
    \caption{Family of textures obtained from the twisted meron tube shown in panel (a), by following route (i), \textit{adding modulations}, with the panels differing in the number of in-plane modulations.
    }
\label{fig:meron_inplane}
\end{figure}

Following route (iii), \textit{entangling textures}, a continuous twist of the meron helicity along the tube axis generates the twisted meron tube, or simple screw dislocation, shown in Fig.~\ref{fig:meron_inplane}(a), as previously predicted in Ref.~\cite{Azhar2022}. 
In the following, we consider the twisted meron tube as the parent configuration for constructing a new set of three-dimensional textures.

Following route (i), \textit{adding transverse modulations}, the vorticity of the twisted meron tube can be changed, resulting in the textures shown in Fig.~\ref{fig:meron_inplane}(b) and (c).

\begin{figure}
    \centering
    \includegraphics[width=\linewidth]{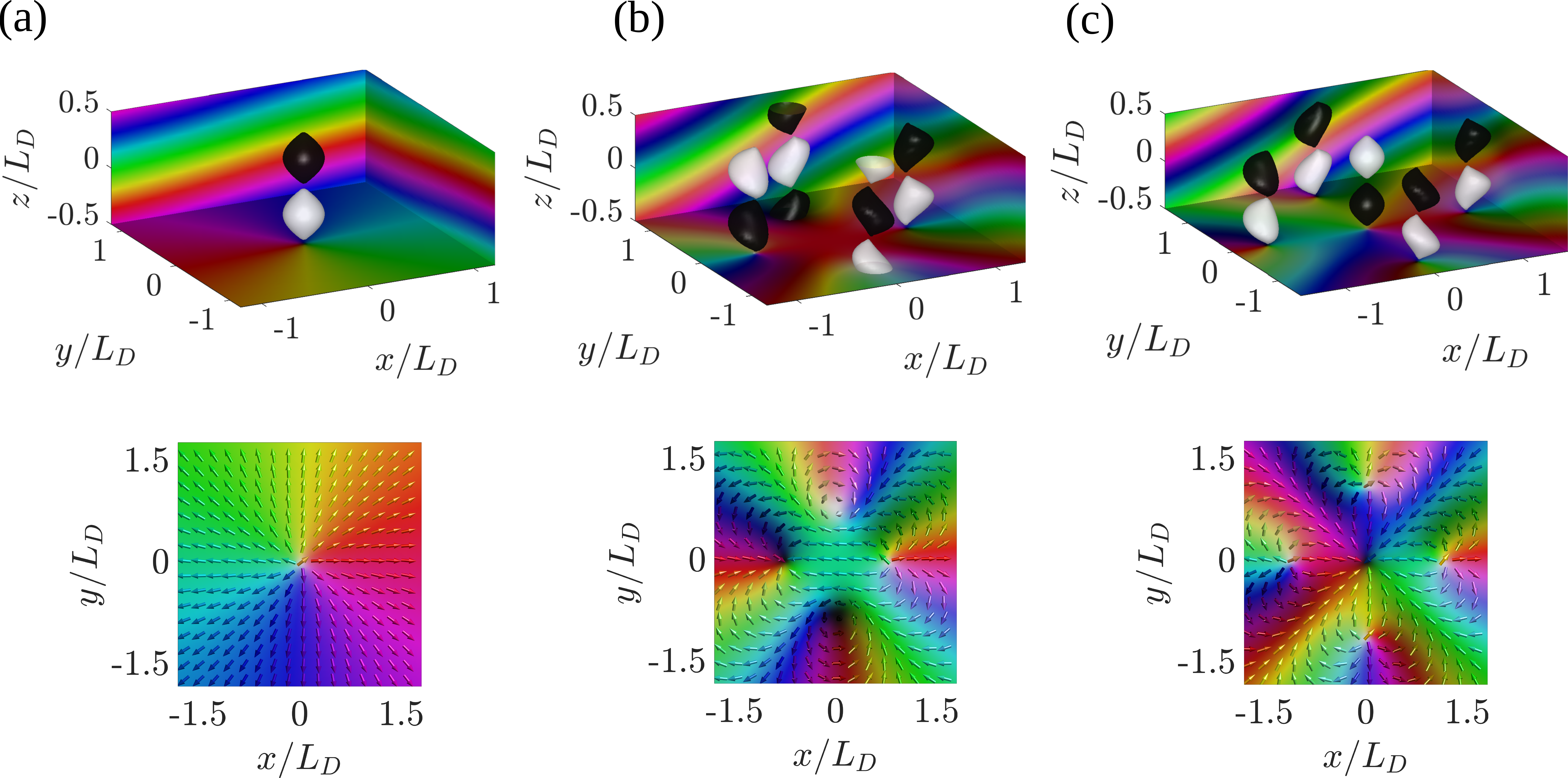}
    \caption{Family of magnetic textures derived from the twisted meron tube by following route (iv), \textit{adding singularities}, and route (ii), \textit{adding structures}. From panel (a) to (c), the number of singularities increases.}
    \label{fig:bloch_points}
\end{figure}

A further family of textures can be obtained by introducing Bloch points into twisted meron tubes. Following route (iv), \textit{introducing singularities}, can yield modulated twisted meron tubes as shown in Fig.~\ref{fig:bloch_points} with potentially further textures added via route (ii).

\subsection{Composite skyrmion–meron textures}
\label{sec:composite}
\begin{figure*}
    \centering
    \includegraphics[width=0.75\linewidth]{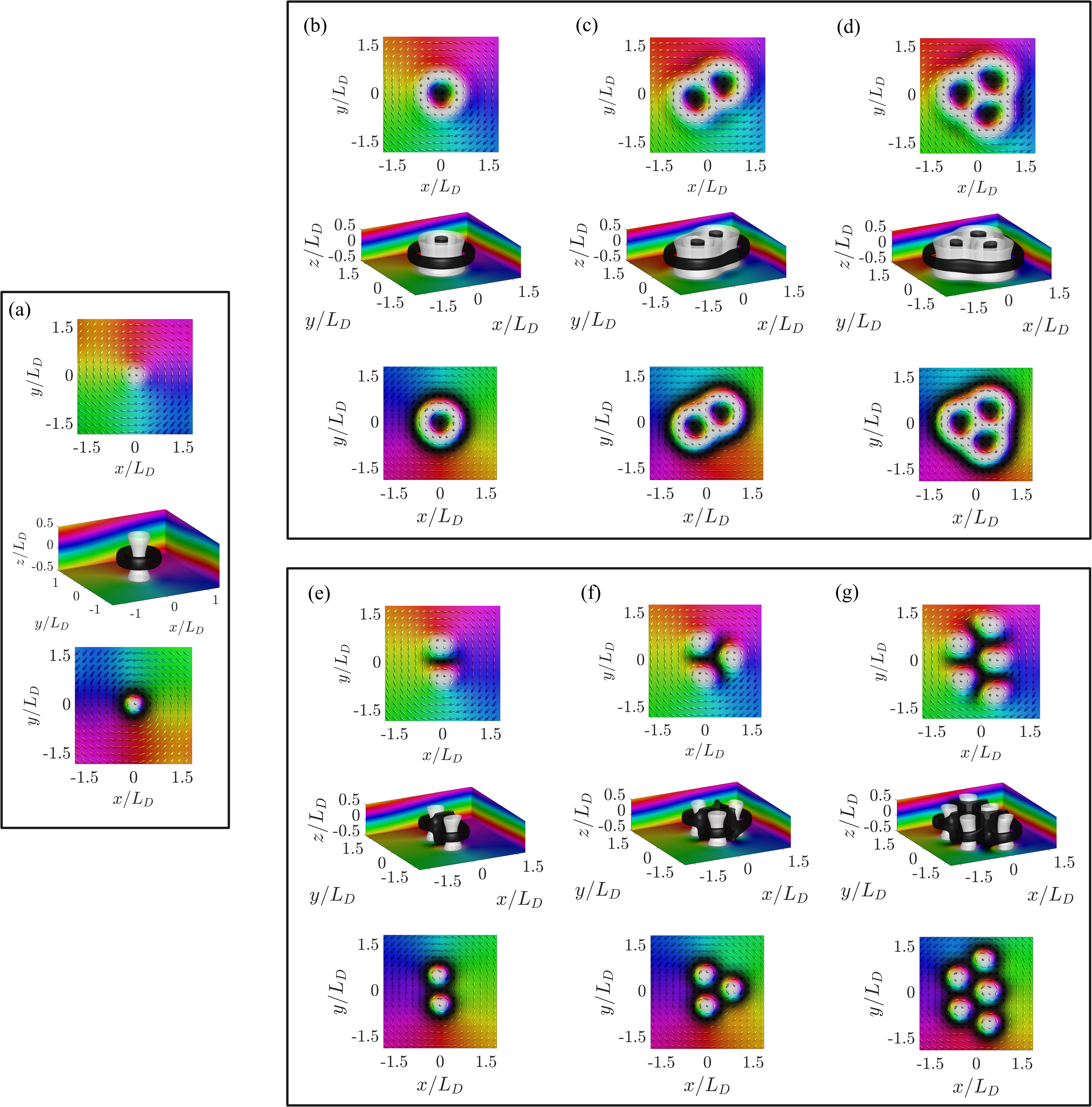}
    \caption{Family of twiston configurations with additional skyrmion tubes introduced either in the central region (upper box) or in the surrounding region (lower box). In both cases, the number of skyrmion tubes increases from left to right. For each configuration, the upper row shows the $xy$ cross-section at $z=0.5$, the middle row shows the three-dimensional magnetization texture, and the lower row shows the $xy$ cross-section at $z=0$.
    }
    \label{fig:half_hopfion}
\end{figure*}

Shown in Fig.~\ref{fig:half_hopfion}(a) is the twiston, a meron surrounded by a meron tube as discussed in Sec.~\ref{sec:general}\ref{sec:meron}.

Adding skyrmions at the center of the twiston, with polarity opposite to that of its core, yields the family of textures shown in Fig.~\ref{fig:half_hopfion}(b)--(d), where the number of added skyrmions increases from (b) to (d), similar to those reported in Ref.~\cite{Zheng2023}.

Adding skyrmions with the same polarity as the twiston core yields the textures in Fig.~\ref{fig:half_hopfion}(e)--(g), with their number increasing from left to right. The interpretation in terms of added skyrmion tubes is only locally valid, as illustrated in the lowest row of each panel. Horizontal cross-sections through the upper and lower regions of the three-dimensional magnetization texture instead reveal fractionalization into vortex and antivortex constituents of opposite polarities, as exemplified in the top row of each panel. These textures are also referred to as skyrmion bundles~\cite{tang2021,zhang2024}.

\begin{figure}
    \centering
\includegraphics[width=\linewidth]{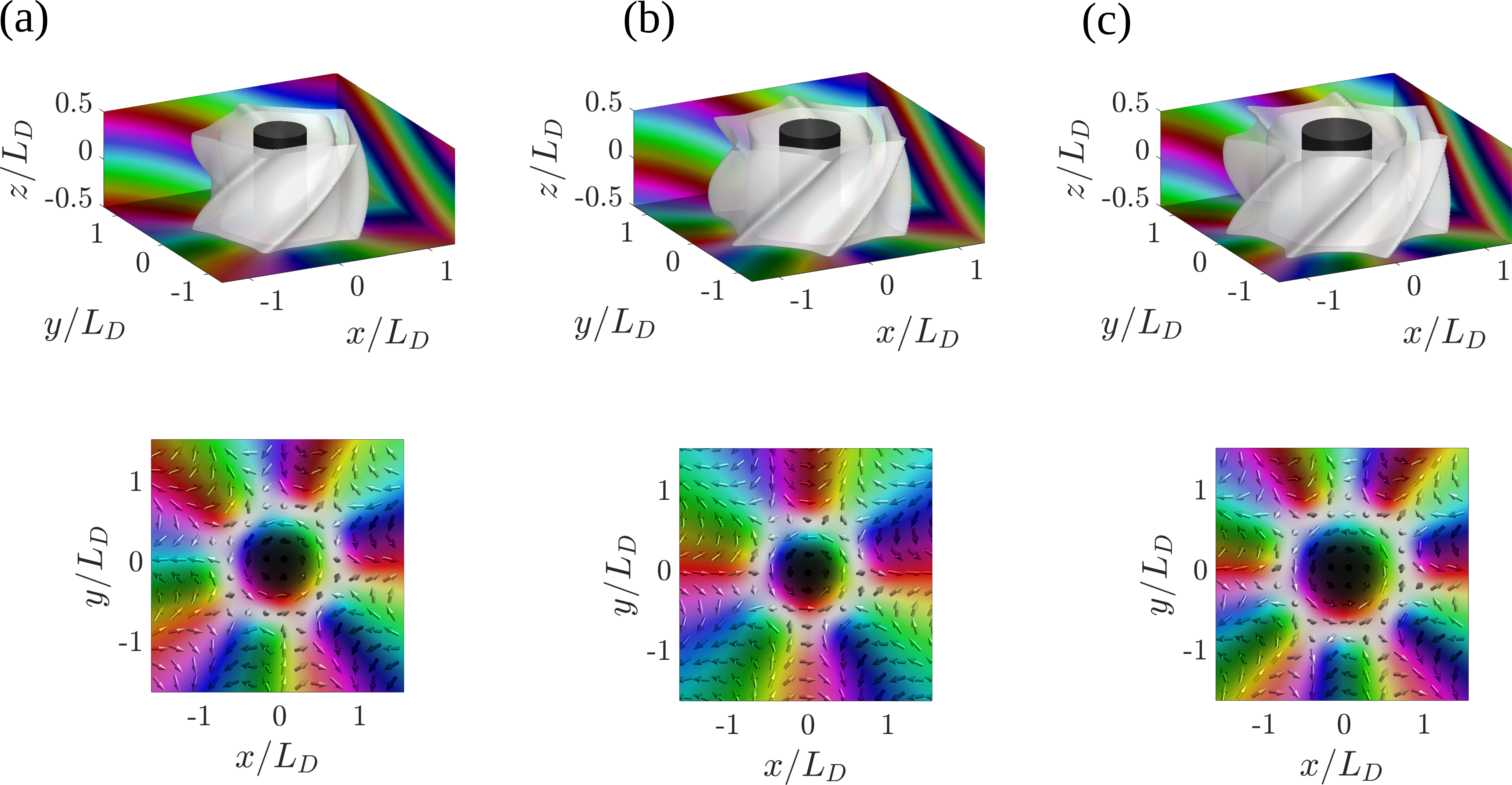}
    \caption{Family of magnetic textures derived from the parent meron tube by combining route (i), \textit{adding modulations}, route (ii), \textit{adding structures}, and route (iii), \textit{entangling textures}. The textures in the different panels are distinguished by the number of in-plane modulations introduced into the outer meron tube. 
    }
    \label{fig:combined_texture}
\end{figure}

The family of textures shown in Fig.~\ref{fig:combined_texture} is constructed by combining routes (i)–(iii). Starting from the parent skyrmion tube , route (ii), \textit{adding structures}, is first applied by introducing an additional meron tube of opposite polarity surrounding the parent texture. Route (i), \textit{adding modulations}, is then applied to the outer meron tube by introducing different numbers of in-plane modulations, which distinguish panels (a)–(c).  Finally, through route (iii), \textit{entangling textures}, the modulated outer meron tube is twisted along the $z$-direction, giving rise to the family of newly predicted screw dislocations shown in the figure.

The construction principles considered here act directly on the geometry and internal structure of a parent texture and are sufficiently general to encompass the textures generated by other design strategies~\footnote{The textures shown in Figs.~\ref{fig:braided_textures},\ref{fig:meron_inplane},\ref{fig:half_hopfion},\ref{fig:combined_texture} are also reported in the PhD thesis of Sandra Chulliparambil Shaju~\cite{Shaju2026Thesis}}. 
For example, textures generated by ``spiralization'' or ``despiralization'' can still be understood within the same construction principles~\cite{Shaju2026}.

\subsection{Metastability in a representative bulk chiral-magnet model}
By means of micromagnetic simulations, we verified that all configurations reported in Sec.~\ref{sec:general}\ref{sec:meron},~\ref{sec:composite} are metastable within the bulk chiral magnet model
\begin{equation}
    E=\int \mathrm{d}^3r \, [A\, (\vect{\nabla} \vect{m})^2 \:+D \,\vect{m}\cdot\boldsymbol{\nabla}\times\vect{m}
    ],
\label{eq:chiralMagnets}
\end{equation}
where $A$ is the ferromagnetic exchange stiffness, $D$ is the Dzyaloshinskii-Moriya interaction (DMI) strength, and $\vect{m}$ is the normalized magnetization vector. Details about the simulation are given in App.~\ref{app:micromag}.

\section{Topological Design Rules based on Construction Routes}

In this section, we discuss how the topology of the resulting three-dimensional textures changes relative to their parent configurations for each construction route. To characterize these changes, we use the discrete Hopf index formula introduced in Ref.~\cite{Azhar2026}. Within this formalism, a three-dimensional magnetic texture is decomposed into $N_\Phi$ flux tubes of the emergent magnetic field $\vect{F}$, defined as
\begin{equation}
F^\gamma=\frac{1}{8\pi}\epsilon^{\alpha \beta \gamma}\,\vect{m}\cdot\left(\partial_\alpha\vect{m}\times\partial_\beta\vect{m}\right),
\end{equation}
where $\epsilon^{\alpha \beta \gamma}$ denotes the Levi-Civita symbol and $\alpha, \beta, \gamma\in{x,y,z}$. Each flux tube $i$ carries a flux $\Phi_i=\left|\int_{\vect{a}_i}\mathrm{d}\vect{a}\cdot\vect{F}\right|$, where $\vect{a}_i$ denotes a cross-sectional area of the corresponding flux tube. In this representation, the Hopf index can be expressed as~\cite{Azhar2026}
\begin{equation}
H=\sum_{i=1}^{N_\Phi} L_{ii}\Phi_i^2
+ \sum_{\substack{i,j=1\\ j\neq i}}^{N_\Phi} L_{ij}\Phi_i\Phi_j.
\label{eq:Hopf_decomposed}
\end{equation}
Here, $L_{ii}$ is the oriented self-linking number of flux tube $i$, while $L_{ij}$ denotes the oriented linking number between flux tubes $i$ and $j$, i.e., their inter-linking.

In the following, $H_0$ represents the Hopf index per period of the parent configuration under consideration. We now consider each of the construction routes individually and examine how the topology of the resulting textures is related to that of their parent configurations.

\subsubsection{Adding transverse modulations}
Introducing additional in-plane modulations into a parent structure changes its vorticity $V$ and produces a new period that scales as $1/V$. For a fixed volume, the Hopf index scales quadratically with $V$, whereas over one period of the resulting texture it scales as $H \sim V H_0$.

\subsubsection{Adding structures}
When adding a structure, one must distinguish whether new inter-linkings are introduced. In their absence, adding a structure with Hopf index $H_1$ to a parent texture with Hopf index $H_0$ gives $H = H_0 + H_1$, as described by the first term of Eq.~\ref{eq:Hopf_decomposed}. If new inter-linkings are introduced, the second term provides additional contributions to the total Hopf index.
For example, the families in Fig.~\ref{fig:half_hopfion}(b)--(g) are derived from the twiston shown in panel (a), which has Hopf index $H_0=3/4$. Adding $N$ skyrmions yields $H=H_0-N$ for the textures in panels (b)--(d). Reversing the polarity of the added skyrmions reverses the sign of the inter-linking contribution, giving $H=H_0+N$ for the textures in panels (e)--(g).

\subsubsection{Entangling textures}
Entangling textures typically modifies either their self-linking or their inter-linking.

Braiding two parent textures with Hopf indices $H_0^{(1)}$ and $H_0^{(2)}$ produces a composite structure with inter-linked flux tubes. If each parent is considered as a single flux tube carrying flux $\Phi_1$ or $\Phi_2$, respectively, 
and braiding preserves their self-linking, the resulting Hopf index is
$H=H_0^{(1)}+H_0^{(2)}+2 L_{12}\Phi_1\Phi_2,$
where $L_{12}$ denotes the signed inter-linking number of the two tubes. All Hopf indices and linking numbers are evaluated over the same spatial period.

When a twist (helicity rotation along the longitudinal direction of the parent texture) is introduced, the self-linking number of the corresponding texture is modified. For a full twist, the self-linking changes by 1.

For textures with an $n$-fold rotational symmetry, also fractional braids and twists by an angle $360^\circ/n$. 
 are allowed in the same volume, thereby enlarging the periodicity of the new texture. 

Closing a flux tube, for example, by joining the ends of a twisted skyrmion tube to form a hopfion, leaves the Hopf index unchanged, provided that its flux, self-linking, and interlinking with other tubes are preserved upon closing the tube.

\subsubsection{Introducing singularities}
In the presence of a Bloch point, the magnitude of the magnetization vanishes locally. As a result, the normalized magnetization field becomes ill-defined at the Bloch point, and the usual homotopy-based classification is no longer directly applicable. The topological classification of such textures is discussed in Ref.~\cite{Gessler2026}.

\section{Conclusion}
In this work, we introduced a systematic framework for constructing three-dimensional magnetic textures from simpler parent configurations. The framework comprises four elementary construction principles: adding transverse modulations, adding structures, entangling textures, and introducing singularities. Applied individually or in combination, these principles provide a unified perspective on the relationships between structurally distinct three-dimensional magnetic textures. 

Using the skyrmion tube as a parent configuration, we showed that a broad range of previously reported textures, including higher-order skyrmionic structures, twisted and braided skyrmion tubes, hopfions, and Bloch-point-containing textures, can be understood within this common framework. 

Applying the same construction principles to meron tubes allowed us to systematically generate new families of three-dimensional magnetic textures. 
Micromagnetic simulations demonstrate that the predicted configurations can exist as metastable states, with bulk chiral magnets providing one example.

We further related the construction principles to changes in topology using the Hopf index, showing how transverse modulations, added structures, and entanglement modify the topological properties, while singularities require a different kind of classification.

In summary, our framework thus unifies the construction and classification of three-dimensional magnetic textures, connects structural transformations to topological changes, and provides a basis for the systematic exploration of new magnetic states.
\vskip6pt

AI-assisted language editing was used.
%\ethics{}
The ovf data files for the new textures reported in this manuscript (Sec.~\ref{sec:general}\ref{sec:meron} and \ref{sec:composite}) will be made publicly available on Zenodo upon acceptance of the manuscript.
We thank the TWIST team and Markus Garst for fruitful discussions, and Markus Gansel for assistance with the graphics in Fig.~\ref{fig:summary_routes}.
This work was supported by the German Research Foundation (DFG) under Project No.~505561633 as part of the TOROID project, co-funded by the French National Research Agency (ANR) under Contract No.~ANR-22-CE92-0032. Furthermore, we acknowledge funding from the DFG through Project No.~278162697 (SFB~1242, project~B10) and the Volkswagen Foundation~-~0202848. M.~A.~acknowledges funding from the TORUS project within the UDE Postdoc Seed Funding program. S.~C.~S.~acknowledges support from the Studienstiftung des Deutschen Volkes.

\bibliography{references}
\section{Appendix}
\label{app:micromag}
We perform micromagnetic simulation using \texttt{MuMax3}~\cite{Vansteenkiste2014} with the following parameters: exchange stiffness $A = \SI{4e-13}{\joule\per\meter}$, bulk DMI constant $D = \SI{2.8e-4}{\joule\per\square\meter}$, and saturation magnetization $M_{\mathrm{s}} = \SI{1.63e5}{\ampere\per\meter}$. The grid size was set to $(N_x,N_y,N_z) = (180,180,36)$, with a cell size of $\SI{0.5}{\nano\meter}$. Periodic boundary conditions were applied along the $z$-axis. Using these parameters, we obtain $L_{\mathrm{D}}=4\pi A/D=\SI{18}{\nano\meter}$ as the period of the spiral along the $z$-direction.

\end{document}